\documentclass[conference]{IEEEtran}
\IEEEoverridecommandlockouts
\usepackage{xcolor}
\usepackage{amssymb}
\usepackage{siunitx}
\usepackage[hidelinks]{hyperref}
\usepackage{subcaption}
\usepackage[cmex10]{amsmath}
\usepackage[ruled]{algorithm2e}
\usepackage[shortlabels]{enumitem}

\usepackage[style=ieee,backend=bibtex,url=false,doi=false,isbn=false,date=year,citestyle=numeric-comp,maxbibnames=10,minbibnames=10,maxcitenames=10,mincitenames=10]{biblatex}
\bibliography{Aggregation.bib,Convex_models.bib,alexLit}
\usepackage{graphicx} % for pdf, bitmapped graphics files
\usepackage{amsmath} % assumes amsmath package installed
\title{
Complexity Reduction for TSO-DSO Coordination: \\
Flexibility Aggregation vs. Distributed Optimization
\thanks{MBB acknowledges financial support by the German Federal Ministry for Economic Affairs and Climate Action (BMWK) under agreement no. 03EI4043A~(Redispatch 3.0).

TF acknowledge support in the course of TRR 391 Spatio-temporal Statistics for the Transition of Energy and Transport (520388526) by the Deutsche Forschungsgemeinschaft (DFG, German Research Foundation).}
}
\SetNlSty{textbf}{}{.}

\begin{document}

%\author{Maísa Beraldo Bandeira, Alexander Engelmann, and Timm Faulwasser
%\thanks{MBB is with the Institute of Energy Systems, Energy Efficiency and Energy Economics, TU Dortmund University, Dortmund, Germany,  {\tt\small maisa.bandeira@tu-dortmund.de}. This research was conducted while AE was with the same institution as MBB {\tt\small alexander.engelmann@ieee.org}. TF is with the Institute of Control Systems, Hamburg University of Technology, Hamburg, Germany.  \tt\small timm.faulwasser@ieee.org}
%\thanks{MBB acknowledges financial support by the German Federal Ministry for Economic Affairs and Climate Action (BMWK) under agreement no. 03EI4043A~(Redispatch 3.0).}%}

\author{\IEEEauthorblockN{1\textsuperscript{st} Maísa Beraldo Bandeira}
\IEEEauthorblockA{\textit{ie3} \\
\textit{TU Dortmund University}\\
Dortmund, Germany \\
maisa.bandeira@tu-dortmund.de}
\and
\IEEEauthorblockN{2\textsuperscript{nd} Alexander Engelmann}
\IEEEauthorblockA{\textit{logarithmo GmbH \& Co. KG} \\
Dortmund, Germany \\
alexander.engelmann@ieee.org}
\and
\IEEEauthorblockN{3\textsuperscript{rd} Timm Faulwasser}
\IEEEauthorblockA{\textit{Institute of Control Systems} \\
\textit{Hamburg University of Technology}\\
Hamburg, Germany \\
timm.faulwasser@ieee.org}
}

\thispagestyle{empty}
\pagestyle{empty}

\maketitle
%%%%%%%%%%%%%%%%%%%%%%%%%%%%%%%%%%%%%%%%%%%%%%%%%%%%%%%%%%%%%%%%%%%%%%%%%%%%%%%%
\begin{abstract}
The increasing number of flexible devices and distributed energy resources in power grids renders the coordination of transmission and distribution systems increasingly complex. In this paper, we discuss and compare two different approaches to optimization-based complexity reduction: Flexibility aggregation via Approximate Dynamic Programming (ADP) and distributed optimization via the Alternating Direction Method of Multipliers (ADMM). 
Flexibility aggregation achieves near-optimal solutions with minimal communication. However, its performance depends on the quality of the approximation used. In contrast, ADMM attains results closer to the centralized solution but requires significantly more communication steps.
We draw upon a case study combining different \textrm{matpower} benchmarks to compare both methods. 
\end{abstract}

%\Ae{Alternative Subtitle: Aggregation vs. Distributed Optimization? Talk about aggregation via ADP in the abstract and intro? I think this is closer to power system wording.}

%%%%%%%%%%%%%%%%%%%%%%%%%%%%%%%%%%%%%%%%%%%%%%%%%%%%%%%%%%%%%%%%%%%%%%%%%%%%%%%%

%\hspace{-5cm}
\section{Introduction} \label{sec:Intro}
%\Ae{Feel free to put me wherever in the order of authors.}
%The growing number of Distributed Energy Resources (DER) in the power grid renders the coordination of power exchange between voltage levels increasingly complex. In classic settings, only  transmission system operation relies on numerical optimization.

Due to the increasing number of flexible devices in  distribution grids, the operational principles for  Distribution System Operators (DSOs) have to be rethought. In this context, optimization-based solutions exploiting the hierarchical structure of power systems play a key role~\cite{Molzahn2017}. 
Transmission and distribution grids are physically connected but owned and operated by different entities, i.e., Transmission System Operators (TSOs) and DSOs. Hence, finding a feasible or even optimal solution to power dispatch requires coordination between them. There are two main approaches in the literature to exploit the grid structure for decentralized or distributed coordination and operation: \textit{distributed optimization} and \textit{flexibility aggregation}.
The former partitions the overall optimization problem into smaller but coupled subproblems. These subproblems are then solved using distributed or decentralized numerical optimization algorithms. Examples for such algorithms include the Alternating Direction Method of Multipliers (ADMM)– \cite{Erseghe2014,Kim2000,boyd2011}, the Augmented Lagrangian Alternating Direction Inexact Newton (ALADIN) method~\cite{Houska2016,Dai2022, Engelmann2019c} and other methods such as, e.g.,  \cite{Kim1997,Hug-Glanzmann2009,Engelmann2021a}. We refer to~\cite{Mohammadi2019,Li2018a,dai2024real} for works addressing TSO-DSO coordination. 

%\tf{adjust references as needed} \MB{@Alex: could you take a look at the dopt citations?}
%\MB{to find: distributed algorithm applied to TSO DSO coordination specifically? Maybe not relevant?}
%\Ae{Checkout \cite{Mohammadi2019,Li2018a} + refs therein.}

The latter approach, flexibility aggregation, captures the combined flexibility of all devices in the distribution grid at the point of interconnection with the transmission grid.
Several approaches exist to estimate or compute the flexibility at the vertical interconnection to the upper-level grid ~\cite{MayorgaGonzalez2018,sarstedt2022,silva2018}. 
%With the exception of~\cite{majumdar2023}, all works assume a unique coupling point. 
A commonality of these approaches is that they compute an approximation of the set of admissible operating conditions of active and reactive power called \emph{Feasible Operating Region (FOR)} via sampling. This is either based on massive sampling of set-point changes of all flexible devices~\cite{MayorgaGonzalez2018,sarstedt2022}, or based on optimization~\cite{silva2018}, which try to explore the boundaries more systematically.
Based on these samples, one approximates the FOR, e.g., by taking their convex hull. 
%and, therefore, to the infeasibility of the disaggregation problem.
In general, providing guarantees on the feasibility of aggregation-disaggregation schemes while considering grid constraints remains an open problem that is not fully tackled in the available literature. An exception is \cite{nazir2022}, which derives nodal injection boundaries for radial distribution grids using an inner approximation of an optimal power flow problem.
Moreover, many aggregation schemes consider the voltage at the grid interconnection to be constant (thus they only summarize the power flexibility at the interconnection), exceptions are~\cite{sarstedt2020, meinecke2023} and the cost of flexibility usage is rarely considered,  see, e.g., 
%Besides in \cite{silva2018,sarstedt2022},  most works% presumably because of its inherently non-differentiable and non-convex nature.
~\cite{silva2018,sarstedt2022, Capitanescu2023}.

Recently, we proposed to tackle flexibility aggregation  via Approximate Dynamic Programming (ADP)~\cite{Engelmann2025,Bandeira2024}, i.e., a hierarchical approach that leverages the tree structure of the network consisting of a transmission grid and multiple distribution grids.
%\Ae{The last two sentences seem to be redundant.}
From the ADP perspective, aggregation is a set projection of the constraint sets of the DSOs onto their coupling variables to the TSO. For affine grid models such as the DC model, this projection can be obtained via tools from computational geometry \cite{jones2004} while avoiding any sampling. 

%While TSO-DSO coordination is crucial in power system operation, we were not able to find an direct comparison of aggregation approaches with distributed optimization. 
The present paper discusses and compares TSO-DSO coordination based on distributed optimization with our approach for flexibility aggregation via ADP. We explore advantages and drawbacks regarding applicability, computational, and communication requirements compared to the centralized solution. Additionally, we explore both approaches regarding optimality and feasibility guarantees.

The remainder of the paper is structured as follows:
Section~\ref{sec:ProblemStatement} presents the  problem statement, recap distributed optimization approaches, and the ADP approach. Section~\ref{sec:Results} compares and discusses numerical results of both methods. The paper concludes with Section~\ref{sec:conclusion}.

\section{Problem Statement} \label{sec:ProblemStatement}

%\begin{figure}[t]
%	\centering
	%	\hspace{-5em}
%	\includegraphics[trim={5cm 0 0 0},clip, width=1.1\linewidth]{Figures/tso_dso.pdf}
%	\caption{Hierarchically structured power system.}
%	\label{fig:tso_dso}
%\end{figure}
 %Here we briefly introduce the joint 
Power system operation often rely on the solution of Optimal Power Flow (OPF) problems. 
Examples %for such operational problems 
are the re-dispatch of generator set-points due to system overload %\cite{linnemann2011a}
or optimal reactive power dispatch. %\cite{gabash2012}.
At the same time, the power grid is usually divided in transmission and distribution grids operated by TSOs and DSOs.
%TSO-DSO optimization problem, which is described as the Optimal Power Flow (OPF) problem
%Due to the above grid structure,  
This leads to an inherent tree structure, where the TSO is the root of the tree and the DSOs are its leaves. If DSOs are coupled, they can be treated as a single, larger DSO while preserving the tree structure. %, cf.  Figure \ref{fig:tso_dso}.
Coupling between the TSO and DSOs happens via the grid constraints, i.e. nonlinear power flow equations.

Mathematically, one can formalize this tree structure  by introducing subsystems $i \in \mathcal S=\{1,\dots,|S|\}$, where $i = 1$ refers to the TSO and the index set 
$\mathcal D \doteq \mathcal S\setminus \{1\}$ refers to the DSOs. 
Hence, the corresponding OPF problems can be written in the form of
\begin{subequations}\label{eq:prob1} 
	\begin{align} 
		\min_{\{y_i\}_{i \in \mathcal S},\{z_i\}_{i\in \mathcal D}} &f_1(y_1)+ \sum_{i \in \mathcal D} f_i(y_i,z_i)  \\
		\text{ subject to  } & (y_i,  z_i)\in \mathcal{{ X}}_i \text{ for all }i \in \mathcal D \label{eq:locConstr}\\
		&(y_1,z_2,\dots,z_{|\mathcal S|}) \in \mathcal X_1,
	\end{align}
\end{subequations} where $y_i$ are local variables of each system operator, $z_i$ are the coupling variables, i.e., variables that appear both in the TSO and the DSO problem and $\mathcal X_i$ are local constraint sets.  
These sets are described by the nonlinear AC power flow equations for a balanced grid with zero line charging capacities, generator bounds on active and reactive power, voltage bounds, line limits, and renewable generators' capability curves for each subsystem. 
We omit a detailed description of the constraints here and refer to \cite{Bandeira2024} for details on the modeling.

It is difficult to solve \eqref{eq:prob1} across multiple system operators due to its complexity and data confidentiality concerns.
Next, we recall two very different approaches for handling this complexity.

\subsection{Approximate Dynamic Programming}
Feasibility Preserving Approximate Dynamic Programming (FP-ADP) \cite{Engelmann2025,Bandeira2024}
 is based on a crucial observation: in power systems operation, it is especially important to compute feasible solutions while optimality is typically of secondary importance \cite{alizadeh2023}.

%\Ae{I have a (very rough, probably buggy) proposal for the whole theory stuff connecting ADP and ADMM (feel free to modify or ignore it if you think it is not that good).\newline
%1) Start with the problem formulation. Refer to PSCC paper for details on what $\mathcal X_i$ etc is exactly (although describing it in words, i.e. AC OPF with such and such special constraints).  Say that the DSO TSO problem is tree structured and can thus be formulated as the following problem (where SS 1 refers to the TSO):
%\begin{subequations}\label{eq:prob1} 
%	\begin{align} 
%		\min_{\{y_j\}_{j \in \mathcal S},\{z_j\}_{j\in \mathcal D}} f_i(y_1)+& \sum_{i \in \mathcal S \setminus \{1\}} f_i(y_i,z_i)  \\
%		\text{ subject to  } \;  (y_i,  z_i)&\in \mathcal{{ X}}_i \text{ for all }i \in \mathcal D \label{eq:locConstr}\\
%		(y_1,z_2,\dots,z_{|\mathcal S|}) &\in \mathcal X_1
%	\end{align}
%\end{subequations} \newline
%2) Reformulation of \eqref{eq:prob1} via value functions gives the problem formulation from the PSCC paper (I'm not sure whether we should repeat it or not, but we can, maybe you can check where we are in terms of space after the ADMM section, maybe that's also a thing to discuss with Timm). 
%Doing so leads to ADP.
%}

%\subsection{ADP framework}
% We will use some of these tools in the following.

%With these ADP schemes, feasibility of the overall algorithm consisting of aggregation--solving the TSO problem and disaggregation  can be mathematically guaranteed.
%Moreover, it comes with the possibility of multi-level aggregation schemes.

The key idea of FP-ADP is to \textit{hide} the complexity of the DSO problems via value-functions $V_i$, which expresses the aggregated cost of flexibility usage inside of the FOR as a function of the coupling variables $z_i$ for each DSO  $i \in \mathcal D$.
Hence, problem~\eqref{eq:prob1} can be written as
\begin{subequations}\label{eq:TSOProbADP} 
		\begin{align} 
\min_{y_1,\{z_i\}_{i \in \mathcal D}}  &f_1(y_1) + \sum_{i \in \mathcal D} V_i(z_i) \\
	\text{ subject to  } &  (y_1,z_2,\dots,z_{|\mathcal S|}) \in \mathcal X_1, \label{eq:locConstr}
\end{align} 
\end{subequations} 
%\Ae{@Maisa: I have changed the $\bar X_i$ to $\mathcal X_1$ here, I don't think we need $\bar X_1$ here yet and I think this way it might be easier to get for teh reader.}
%which depend on the coupling variables $z_i$ but not directly on the local TSO variables $y_i$. The value functions $V_i$ are calculated via  \textit{DSO problems}
where the DSO value functions are given by
\begin{subequations}\label{eq:DSOprobs} 
	\begin{align} 
		V_i(z_i)\doteq \min_{y_i} \,\,  &f_i(y_i,z_i)  \\
		\text{ subject to  } &  (y_1, z_i)\in \mathcal{{ X}}_i. \label{eq:locConstr}
	\end{align}
\end{subequations} 
%for all DSOs $i\in \mathcal S \setminus \{1\}$.
Observe that problem~\eqref{eq:TSOProbADP} is equivalent to \eqref{eq:prob1}.
Moreover, notice that \eqref{eq:TSOProbADP} entails fewer decision variables compared to \eqref{eq:prob1}.
Thus, if the DSOs are able to compute easy-to-handle expressions for the functions $V_i$, complexity reduction can simply happen by communicating all $V_i$s to the TSO, which in turn solves the low-dimensional problem~\eqref{eq:TSOProbADP}.
Unfortunately, computing the value functions $V_i$ is often difficult.

Feasibility of the DSO problems~\eqref{eq:DSOprobs} is characterized by the domain of $V_i$, i.e. $\operatorname{dom}(V_i)\doteq \{z_i\;|\; V_i(z_i) < \infty\}$.
The key observation of FP-ADP is that one can compute $\operatorname{dom}(V_i)$ by a set projection of the local constraint sets $\mathcal X_i$ onto the coupling variables $z_i$ \cite{Engelmann2025}.
This projection is then considered in the TSO problem~\eqref{eq:TSOProbADP}  as an additional constraint.
Sampling-based schemes for the non-convex AC power flow model can approximate this projection. 
They can even be computed exactly in affine grid models such as \texttt{LinDistFlow} or the DC model.

\subsection{Complexity Reduction with FP-ADP}
 Formally, the \emph{set projection} of $\mathcal X \subseteq \mathcal Y \times \mathcal Z$ onto $\mathcal Z$ is \cite{Rakovic2006}
\begin{align}\label{eq:setProj}
	\operatorname{proj}_{\mathcal Z}(\mathcal X) \doteq \{ z \in \mathcal Z  \;|\;\exists \; y \in \mathcal Y \text{ with  } (z,y) \in \mathcal X\} \subseteq \mathcal Z.
\end{align}
It has been shown in \cite[Lemma 1]{Engelmann2025} that \eqref{eq:TSOProbADP} is equivalent to
\begin{subequations}\label{eq:TSOProbADP2} 
	\begin{align} 
		\min_{y_1,\{z_i\}_{i \in \mathcal D}}  &f_1(y_1) + \sum_{i \in \mathcal D} V_i(z_i) \\
		\text{ subject to  }  & (y_1,z_2,\dots,z_{|\mathcal S|}) \in \bar {\mathcal  X}_1, \label{eq:locConstr}
	\end{align} 
\end{subequations} 
where
%Moreover, the modified constraint set $\bar {\mathcal X}_1$ is defined as
%Moreover, the constraint \eqref{eq:locConstr} can be reformulated as
%$\bar {\mathcal X}_j \doteq\mathcal{X}_j(z_j) \cap (\mathbb R^{n_{yj}} \times  (\cap_{i \in \mathcal C}  \mathcal P_i  )$.
\begin{align*}
\displaystyle \bar {\mathcal X}_1 \doteq\mathcal{X}_1 \cap (\mathbb R^{n_{y1}} \times \bigcap_{i \in \mathcal D}   \mathcal P_i).
\end{align*} 
Here, $\mathcal P_i$ is defined as 
\begin{align} \label{eq:proj}
 	\mathcal P_i \doteq \operatorname{proj}_{ \mathcal Z_i}{ {\mathcal  X}}_i \quad i \in \mathcal D,
 \end{align} 
i.e., the projection of the constraint sets of the DSOs $\mathcal X_i$ onto their coupling variables to the TSO $z_i$.

%  the \emph{set projection} of $\mathcal X \subseteq \mathcal Y \times \mathcal Z$ onto $\mathcal Z$ \cite{Rakovic2006}
% \begin{align}\label{eq:setProj}
% 	\operatorname{proj}_{\mathcal Z}(\mathcal X) \doteq \{ z \in \mathcal Z  \;|\;\exists \; y \in \mathcal Y \text{ with  } (z,y) \in \mathcal X\} \subseteq \mathcal Z.
% \end{align}

%
%In the power system setting problem, the problems defined via \eqref{eq:ValFunRef0} are ``local'' OPF problems for all DSOs $i \in \mathcal D$ for the active/reactive power fixed at the interconnection, and $V_1$ is the OPF problem for the TSO.
%\footnote{Note that with slight abuse of notation, we use the same symbols $f_i$ and $\mathcal X_i$ in the definition of $\mathcal X_i$, \eqref{eq:locCost}, and \eqref{eq:ValFunRef0} although these sets/functions depend on their local and coupling variables in \eqref{eq:ValFunRef0} only.}
%Moreover, note that computing the set projection \eqref{eq:proj} is, by the definition of the set projection \eqref{eq:setProj} and the definition of the $z_i$ variables, equivalent to computing all $(p_{k,l}, q_{k,l})$-values at the interconnection transmission lines between TSO and DSOs $(k,l) \in \mathcal B_i^c$  for which the DSO problems are feasible.
%In other words, it is  equivalent to what is frequently called ``aggregation''  in the literature \cite{?}.

%\subsubsection{Feasibility-Preserving ADP}

The solution to problem~\eqref{eq:TSOProbADP2} is calculated in two sweeps. 
In Step 1), the projections $\mathcal P_i$ of the feasible sets of all DSO problems $j \in \mathcal D$ are computed.
Step 2) includes all $(\mathcal P_i, \tilde V_i)$ in the TSO problem~\eqref{eq:TSOProbADP}, which is solved in Step 3).
The optimal coupling variables $\bar z_i^\star$,  are communicated to the DSOs in Step 4). 
The DSOs solve their OPF problems~\eqref{eq:DSOprobs} for fixed $\bar z_i^\star$  in Step 5).
The overall scheme is summarized in \autoref{alg:appDP}.

\begin{algorithm}[t]
	\SetAlgoLined
	\caption{Feasibility-Preserving ADP for \eqref{eq:prob1}. }\label{alg:appDP}
	\BlankLine
	\textbf{Backward sweep:}\\
	%\begin{enumerate}
		1) Compute the FORs  $\mathcal P_i$ via \eqref{eq:proj}, and  $V_i$ for all DSOs $i \in \mathcal C$ and send them to the TSO. \label{step:aggre}\\
		2) Include all  $(V_i,\mathcal P_i)$ into the TSO problem~\eqref{eq:TSOProbADP}. \label{step:comm1} \\
	%\end{enumerate}
	\textbf{Forward sweep:} \\
	%\begin{enumerate}
	%	\setcounter{enumi}{2} % Start counting from 3
		3) Solve  \eqref{eq:TSOProbADP}. \label{step:solveTSO}\\
		4) Distribute optimal coupling variables $\bar z_i^\star$ to all DSOs $i \in \mathcal D$. \label{step:comm2}\\
		5) Solve \eqref{eq:DSOprobs} for fixed $\bar z_i^\star$ at all DSOs. \\\label{step:disaggre}
	%\end{enumerate}
	\BlankLine
\end{algorithm}

%\begin{algorithm}[t]
%	\SetAlgoLined
%	\caption{Feasibility-Preserving ADP for \eqref{eq:prob1}. \Ae{I think the algorithm environment in our ADP paper looks a bit nicer. Adapt}\MB{done}} \label{alg:appDP}
%	\BlankLine
%	%	\textbf{Initialize} $\mathcal H \doteq \mathcal L$ \\
%	\textbf{Backward sweep:}\\
%	1) Compute the FORs  $\mathcal P_i$ via \eqref{eq:proj}, and  $V_i$ for all DSOs $i \in \mathcal C$ and send them to the TSO. \label{step:aggre}\\
%	2) Include all  $(V_i,\mathcal P_i)$ into the TSO problem \eqref{eq:TSOProbADP}. \label{step:comm1}\\
	%\textbf{Forward sweep:} \\
%	3) Solve  \eqref{eq:TSOProbADP}. \label{step:solveTSO}\\
%	4) Distribute optimal coupling variables $\bar z_i^\star$ to all DSOs $i \in \mathcal D$.\label{step:comm2}\\
%	5) Solve \eqref{eq:DSOprobs} for fixed $\bar z_i^\star$ at all DSOs. \label{step:disaggre}\\
	%\BlankLine
%\end{algorithm}

Note that if one can compute exact projections $\mathcal P_i$ and exact value functions $V_i$, \autoref{alg:appDP} {recovers the exact solution of the original OPF problem} \eqref{eq:prob1}.
Moreover, if we replace $V_i$ with approximations $\tilde V_i$ and compute $\mathcal P_i$ by suitable, i.e., not too conservative, inner approximations $\tilde {\mathcal P}_i \subseteq\mathcal P_i$, \autoref{alg:appDP}  {is guaranteed to deliver a feasible but potentially sub-optimal solution} \cite[Corollary 1]{Engelmann2025}.
The feasible sets ${ {\mathcal  X}}_i$ are convex polyhedra for specialized grid models such as linear DC power flow. This allows the use of efficient tools from computational geometry to compute $\mathcal P_i$ \cite{jones2004}.%\cite{jones2004,Fukuda2022,Herceg2013}. 

For an in-depth comparison of different approximation strategies for $\mathcal P_i$ and approaches to approximation $\tilde {{V}}_i$ we refer to \cite{Bandeira2024}.
In general, however, the solutions of \autoref{alg:appDP} are sub-optimal when using approximations $\tilde V_i$.

\subsection{Complexity Reduction with Distributed Optimization}
An alternative way of solving problem~\eqref{eq:prob1} is to apply distributed optimization. Here, we use ADMM because of its simplicity, its popularity in the field, and the possibility of easily highlighting the connections to ADP. Moreover, ADMM exhibits convergence guarantees if~\eqref{eq:prob1} is convex.

To apply ADMM, problem~\eqref{eq:prob1} is reformulated in the consensus form, as in \cite{boyd2011}, via variable copies $\{ z_i^\mathrm \tau\}_{i \in \mathcal D}$, $\{ z_i^\mathrm \delta\}_{i \in \mathcal D}$ and consensus constraints $ z_i =  z_i^\mathrm \tau$, $ z_i =  z_i^\mathrm \delta$. Here, the superscript $\cdot^\tau$ refers to TSO variables, while  $\cdot^\delta$ denotes DSO variables. %\MB{double check this change}
Thus, problem~\eqref{eq:prob1} can be written as
\begin{subequations}\label{eq:probADMM} 
	\begin{align} 
		\min_{\substack{\{y_i\}_{i \in \mathcal S},\{z_i\}_{i\in \mathcal D}, \\ \{\bar z_i^\mathrm \delta\}_{i\in \mathcal D}, \{\bar z_i^\mathrm \tau\}_{i\in \mathcal D}}} &f_i(y_1)+ \sum_{i \in \mathcal D} f_i(y_i,z_i^\mathrm \delta)  \\
		\text{ subject to  }& (y_i,  z_i^\mathrm \delta) \in \mathcal{{ X}}_i \text{ for all }i \in \mathcal D \\
		&(y_1, z_2^\mathrm \tau,\dots, z_{|\mathcal S|}^\mathrm \tau) \in \mathcal X_1 \\
		 &z_i =  z_i^\mathrm \tau  \text{ for all } i \in \mathcal D \label{eq:consConstr1} \\
        &z_i =  z_i^\mathrm \delta \text{ for all } i \in \mathcal D \label{eq:consConstr2} 
	\end{align}
\end{subequations} 
The augmented Lagrangian function of $\eqref{eq:probADMM}$ with respect to the consensus constraints  \eqref{eq:consConstr1}-\eqref{eq:consConstr2} 
is given in \eqref{eq:Lagrangian} (on the top of the next page) using the shorthand 
\[
\mathcal L^\rho \doteq L^\rho \left(\{y_i\}_{i \in \mathcal S},\{z_i, z_i^\mathrm \tau, z_i^\mathrm \delta,\lambda_i^\mathrm \tau,\lambda_i^\mathrm \delta\}_{i\in \mathcal D}\right).
\]
\begin{figure*}
\begin{equation} \label{eq:Lagrangian}
\mathcal L^\rho 
=
%
	%& \\
%&\qquad \qquad\qquad\qquad\qquad \qquad 	\) \doteq \\
	 f_i(y_1)+ \sum_{i \in \mathcal D} f_i(y_i,z_i^\mathrm \delta) + \lambda_i^{\mathrm \delta\top }(z_i^\mathrm \delta-z_i) + \lambda_i^{\mathrm \tau\top }(z_i^\mathrm \tau-z_i)
	%&\qquad \qquad \qquad \qquad \qquad \quad 
 + 	 \frac{\rho}{2}\|z_i^\mathrm \delta- z_i\|_2^2 + 	 \frac{\rho}{2}\|z_i^\mathrm \tau- z_i\|_2^2.
\end{equation}     
\end{figure*}
Minimizing $\mathcal L^\rho$ in a coordinate-descent fashion w.r.t. 
\[(y_1,\dots,y_{|\mathcal S|},z_2^\mathrm \tau,\dots,z_{|\mathcal S|}^\mathrm \tau,z_2^\mathrm \delta,\dots,z_{|\mathcal S|}^\mathrm \delta)\] and with respect to $(z_2,\dots, z_{|\mathcal S|})$ separately  combined with multiplier updates yields the ADMM iterations as shown in Algorithm~\ref{alg:ADMM}, cf. \cite{boyd2011}.%\cite{Bertsekas1997a,boyd2011}.

%Another way to use the TSO-DSO coordination to solve the optimization problem is by using distributed optimization.

Due to the nature of the power flow equations, problem~\eqref{eq:prob1} is nonlinear and usually non-convex. Therefore, there is no general guarantee that Algorithm~\ref{alg:ADMM} converges. However, in practice, ADMM seems to work well for solving OPF problems, see, e.g., \cite{Erseghe2014,Mohammadi2019}. Therefore here we proceed with the AC formulation.

\begin{algorithm}[t]
	\SetAlgoLined
	\caption{ADMM for \eqref{eq:prob1}} \label{alg:ADMM}
	\BlankLine
	\textbf{Initialization:} $\lambda_i^{\mathrm \tau0}=\lambda_i^{\mathrm \delta0}=0$ and ${z}_i^0$\\
    \textbf{Repeat until convergence:}\\
    1) Solution of TSO problem ($i=1$): 
            \begin{align*}
			& \big (\{z_i^{\mathrm \tau \,n+1}\}_{i\in \mathcal D}, y_1^{n+1}\big) =  \underset{(y_1,\{z_i^{\mathrm \tau \,n+1}\}_{i\in \mathcal D}) \in \mathcal X_1}{\textrm{argmin}}
            \mathcal L^\rho
        \end{align*}
        
    2) Solution of all DSO problems $i\in \mathcal D$:
    \begin{align*}
		&\big(z_i^{\mathrm \delta \,n+1}, y_i^{n+1}\big) =   \underset{(y_i,z_i^\mathrm \delta)\in \mathcal X_i}{\textrm{argmin}}\quad\mathcal L^\rho
	    \end{align*}
        
	%    \item Solution of TSO and DSO problems $i\in \mathcal S$: 
   %     \begin{align*}
%			& \big (\{z_i^{\mathrm \tau \,n+1}\}_{i\in \mathcal D}, y_1^{n+1}\big) =  \underset{(y_1,\{z_i^{\mathrm \tau \,n+1}\}_{i\in \mathcal D}) \in \mathcal X_1}{\textrm{argmin}}
   %         \mathcal L^\rho
  %      \end{align*}
  %      \begin{align*}
%		    &\big(z_i^{\mathrm \delta \,n+1}, y_i^{n+1}\big) =   \underset{(y_i,z_i^\mathrm \delta)\in \mathcal X_i}{\textrm{argmin}}\quad\mathcal L^\rho
%	    \end{align*}
	   3) Averaging of coupling variables: 
        \begin{align*}
         %   &{z}_i^{n+1} = \frac{z_i^{\mathrm \tau\,n+1}+{z}_i^{\mathrm \delta\, n+1}}{2} 
         &{z}_i^{n+1} = \frac{1}{2} (z_i^{\mathrm \tau\,n+1}+{z}_i^{\mathrm \delta\, n+1})
        \end{align*}
        
	    4) Multiplier updates for all $i\in \mathcal S\setminus\{1\}$:
        \begin{align*}
	      & \lambda_i^{\mathrm \xi \,n+1} = \lambda_i^{\mathrm \xi \,n} + \rho (z_i^{\mathrm \xi \,n+1} - {z}_i^{n+1}), \quad \xi \in \{\tau, \delta\}
        \end{align*}
  %  \end{enumerate}
	\BlankLine
\end{algorithm}

\section{Numerical Case Study}\label{sec:Results}

We test both approaches for a network wherein one transmission grid is connected to two distribution grids. 
We consider the $9$-bus system data from the \textrm{MATPOWER} dataset \cite{chow1982,schulz1974} for the transmission grid tripling the power capacity of the generators and connecting DSO~$1$ to node $8$ and DSO~$2$ to node 6. Both distribution grids are modified versions of the 15-bus grid from \cite{Battu2016}. For DSO~$1$, the loads on buses $17$, $18$, and $22$ are exchanged by generators with doubled capacity. 
As for DSO~$2$, loads on buses $29$ and $38$ are replaced by generators with the same power capacity.
ADMM considers the AC power flow equations, and the arising NLPs are solved with \textrm{IPOPT}~\cite{Wachter2006}.
%\tf{shall we make code available?}

For ADP, we use \texttt{LinDistFlow}, a linearized version of the \texttt{DistFlow} model \cite{baran1989-0}, maintaining information on reactive power and voltage magnitudes, which are crucial in distribution grids. We simulate the results with two variants, with and without a linear approximation of the losses with the first-order Taylor series \cite{Bandeira2024}, $\tilde{\mathcal P}^{LL}_i$ and $\tilde{\mathcal P}^{LDF}_i$ respectively. We compute both approximations using the \textrm{Polyhedra.jl} toolbox \cite{legat2023polyhedral}. All optimization algorithms are implemented using \textrm{JuMP.jl}~\cite{Lubin2023}. 

To approximate the aggregated cost, we include an estimate for $\tilde{V}_i$ by fitting a quadratic function to sampled points of $V_i$, as presented in \cite{Bandeira2024}.

\begin{figure}[t]
	\centering
%	\hspace{-5em}
	\includegraphics[width=0.8\linewidth]{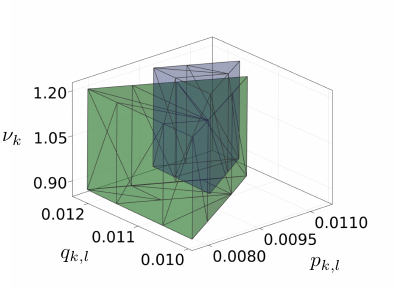}
	\caption{$\mathcal P_i$ for DSO's 1 and 2.}
	\label{fig:3d}
\end{figure}

Independently of the model used in Step~1) of the ADP Algorithm \ref{alg:appDP}, for the disaggregation in Step~5), we consider the AC power flow model. We also add the set point to be attained, from the solution of Step~3), as a soft constraint on Step~5), quadratically penalizing the distance from the set point with a high weight value. The rationale is that the computed set depicted in Figure \ref{fig:3d} is not necessarily an inner approximation of $\mathcal P_i$; it may potentially include infeasible points. In this case, the ADP method does not guarantee feasibility in one shot.

Adhering to the linearized \texttt{DistFlow} model, we consider active and reactive power flow, $p_{k,l}$ and $q_{k,l}$, as well as the squared voltage magnitude $\nu_k$ as coupling variables in the results presented below.

Figure \ref{fig:3d} shows the FORs $\mathcal P_2$ and $\mathcal P_3$ obtained in the ADP scheme using the \texttt{LinDistFlow} model. 
For different values of $\nu$, the $PQ$ area does not change significantly. For easier visualization, we show only the slice $\nu = 1$ of the volume.
%volumes presented in the figure.

\subsubsection{Simulation results} 
The results are shown from the DSOs' point of view, i.e., the power they require from the TSO at the point of interconnection. We simulate the following methods to solve the problem of excess generation in the grid: \\
%\begin{enumerate}
i) Centralized optimization:  Solution of the centralized AC OPF problem.\\
ii) ADMM: Distributed optimization as in Algorithm~\ref{alg:ADMM} to solve the centralized AC OPF.\\
iii) ADP as in Algorithm \ref{alg:appDP} using the AC OPF for Steps~3) and 5), and different approximations for $\mathcal P_i$ and $\mathcal{V}_i$:
\begin{enumerate}[a)]
\item ADP - $\tilde{\mathcal P}^{LL}_i$ : loss linearization formulation to compute the approximation of the FOR and $\tilde{V}_i(z) = 0$.
\item ADP cost -$\tilde{\mathcal P}^{LL}_i$: loss linearization formulation to compute the approximation of the FOR and the quadratic cost approximation from \cite{Bandeira2024}.
\item ADP - $\tilde{\mathcal P}^{LDF}_i$: \textit{LinDistFlow} formulation to compute the approximation of the FOR and $\tilde{V}_i(z) = 0$.
\item ADP cost - $\tilde{\mathcal P}^{LDF}_i$: \textit{LinDistFlow} formulation to compute the approximation of the FOR and the quadratic cost approximation from \cite{Bandeira2024}.
\end{enumerate}

\begin{figure}[htbp]
	\centering
	\begin{subfigure}[b]{0.5\textwidth}
		\centering
	\includegraphics[width=\textwidth]{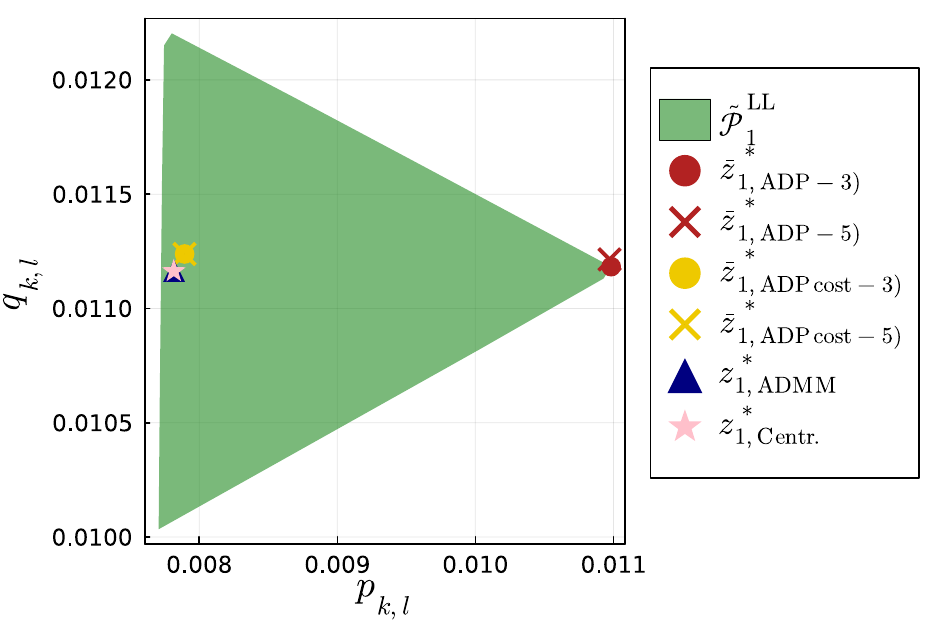}
		\caption{DSO~$1$ interconnection with TSO.}
		\label{fig:z1_LL}
	\end{subfigure}
	\hfill
	\begin{subfigure}[b]{0.5\textwidth}
		\centering
		\includegraphics[width=\textwidth]{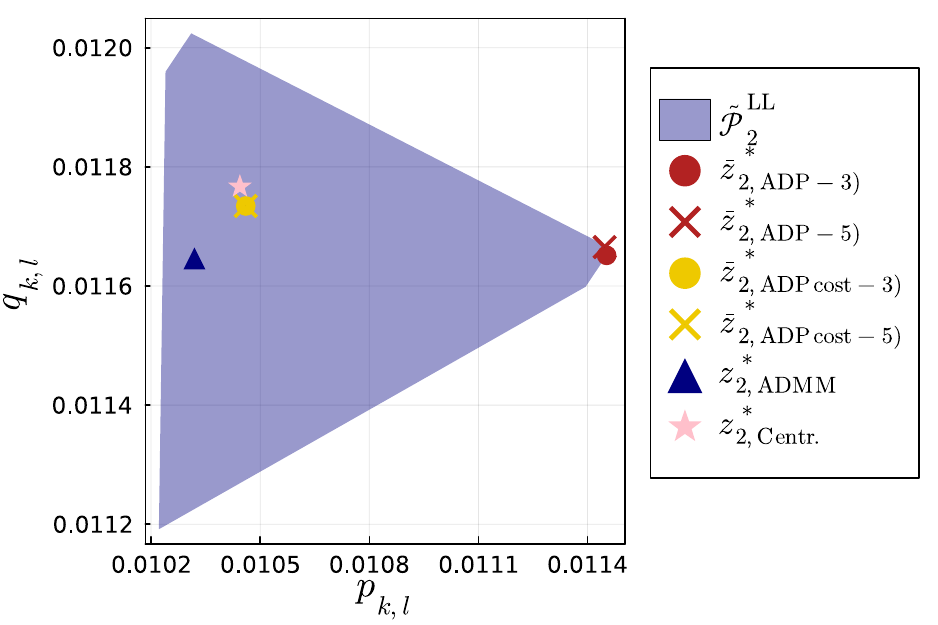}
		\caption{DSO~$2$ interconnection with TSO.}
		\label{fig:z2_LL}
	\end{subfigure}
	\caption{Value of $z_i$ at the interconnection using Loss Linearization formulation for aggregation}
	\label{fig:z_LL}
\end{figure}

%\end{enumerate}
Figure \ref{fig:z_LL} shows the value of the coupling variable for the centralized method and ADMM after convergence, with a required accuracy of $10^{-6}$ as the termination criterion, as well as during the different ADP steps. We use a dot to show Step~3) in Algorithm~\ref{alg:appDP}, i.e. the desired TSO set-point for the coupling variables. A cross shows the final value after disaggregation, Step~5). %The elements in red represent the solutions when no cost is assigned to the FOR ($\tilde{V}_i(z) = 0$), and the values in orange are calculated \MB{using the quadratic cost approximation from \cite{BandeiraPCSS24}.}
For the first simulation results, shown in Figure~\ref{fig:z1_LL}, we use the loss linearization formulation to compute the FOR $\tilde{\mathcal P}^{LL}_i$, which was shown to approximate it closely if the grid is close to the operational point \cite{Bandeira2024}. There is a significant improvement in using a value function, even if it is just an approximation. Otherwise, since there is no cost assigned to the flexibility usage of the DSOs, the TSO optimizer uses as much "free" flexibility from the DSOs as possible. The consequence can be seen in Table \ref{tab:cost}, which shows that the cost of the ADP method without the cost approximation is significantly higher than all other methods. Although the cost of flexible usage with ADMM is the closest to the optimal centralized solution, the cost of ADP using the cost approximation scheme is not much higher, around $1\%$.

\begin{figure}[htbp]
	\centering
	\begin{subfigure}[b]{0.5\textwidth}
		\centering
	\includegraphics[width=\textwidth]{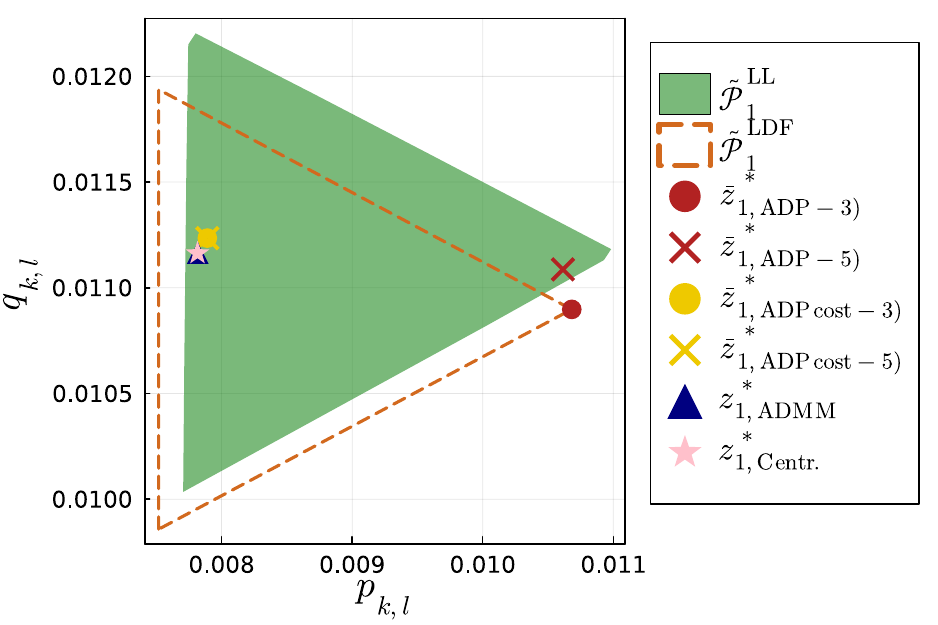}
		\caption{DSO~$1$ interconnection with TSO.}
		\label{fig:z1_LDF}
	\end{subfigure}
	\hfill
	\begin{subfigure}[b]{0.5\textwidth}
		\centering
		\includegraphics[width=\textwidth]{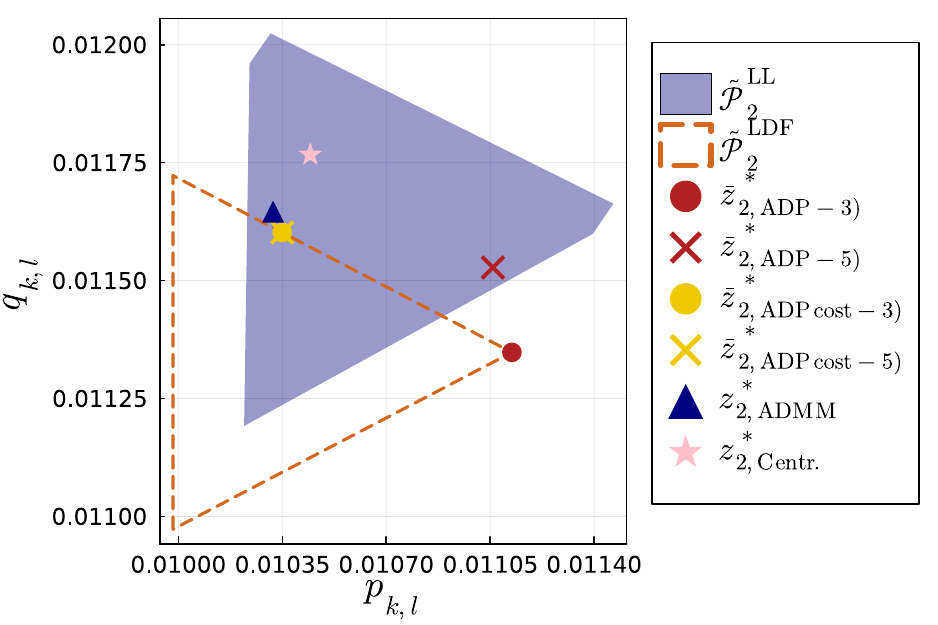}
		\caption{DSO~$2$ interconnection with TSO.}
		\label{fig:z2_LDF}
	\end{subfigure}
	\caption{Value of $z_i$ at the interconnection using \texttt{LinDistFlow} formulation for aggregation}
	\label{fig:z_LDF}
\end{figure}

However, the results using ADP depend on the quality of the FOR approximation. This can be clearly seen when simulating the ADP approach using \texttt{LinDistFlow} for Step~1), see Figure \ref{fig:z_LDF}. In this case, the approximate FOR, $\tilde{\mathcal P}^{LDF}_i$, does not include the optimal solution for the centralized problem. The ADP results do not change for DSO~$1$ when using cost information. Although the solution when no cost information is given is infeasible, the delivered power leads to lower total cost than previously since it requests less power reduction from DSO~$1$. For both DSOs, the optimal point that the TSO computes in Step~3) would lead to infeasibility in Step~5) if the attainment was a hard constraint. Extra steps are then needed. The achievable set point must be communicated back to the TSO, which solves one more optimization step with the actual coupling variable values. 

%The solution to Step~3)\ref{alg:appDP} is shown in red, and the values when considering an approximation of $V_i$ are shown in orange. The optimal solution, i.e., to the centralized problem is shown in pink.

% In fact, we see in Figure \ref{fig:z2} that the set-point $\bar z_2^*$, marked as a red dot, is not feasible and differs slightly from the disaggregated value (crosses).

\begin{table}
\centering
    \caption{Results comparison}
 
\begin{tabular}{l|c|c|c} 
%\hline
\textbf{Algorithm} & \textbf{Total Cost} & \textbf{\# Operations} & \textbf{Comp. Time} \\ 
%\hline

Centr. Opt. & 8.7814 & 1 & 0.2024$\,s$\\ %\hline
ADMM & 8.7818 & 103 & 2.7148$\,s$ \\
$ADP$ - $\tilde{\mathcal P}^{LL}_i$  & 10.7669 & 3 & 1.7200$\,s$ \\ %\hline
$ADP_{cost}$ - $\tilde{\mathcal P}^{LL}_i$  & 8.7928 & 3 & 1.7156$\,s$ \\ %\hlineADP $\tilde{\mathcal P}_i$  & 0.010307 & 0.0114270 & 0.0053225 \\ %\hline
$ADP$ - $\tilde{\mathcal P}^{LDF}_i$ & 10.4012 & 4 & 0.2928$\,s$ \\ %\hline
$ADP_{cost}$ - $\tilde{\mathcal P}^{LDF}_i$  & 8.8361 & 3 & 0.2405$\,s$
\end{tabular} \label{tab:cost}
\end{table}

\subsubsection{Communication and computation effort}
From the practical point of view, the computation burden and the communication effort are critical concerns. Here, communication delays are neglected. However, one of the main advantages of the FP-ADP algorithm is the relatively small number of communication steps. If feasibility can be guaranteed, each DSO communicates with the TSO only twice: back and forth. 

The most expensive computation for FP-ADP is the set projection of $\tilde{\mathcal P_i}$, which took around $1.7\,s$ using the Loss Linearization model and around $0.3\,s$ with \texttt{LinDistFlow}. Including more DSOs should not increase the computation time significantly since each DSO can do this calculation in parallel. Still, the set projection is expected to scale poorly with a larger distribution grid. 

Although the set projection is much faster for the \texttt{LinDistFlow} model due to the smaller number of variables and constraints, the poorer approximation leads to infeasibility in Step~5) when no cost is considered, requiring one extra communication step for the TSO to compute the solution of the new achievable set-point. This could lead to many more back-and-forth optimization and communication steps. 

While for ADMM the overall computation time is not large, around  $2.71\,s$, it can still be improved by parallelizing the computation for the different agents. On the other hand, the algorithm took $103$ iterations to converge with an accuracy of $10^{-6}$ p.u.. With many iterations, the communication time is no longer negligible. Additionally, including more DSOs, might affect the convergence behavior of ADMM. %, but convergence might not be achieved.

%\begin{figure}
%    \centering
%    \includegraphics[trim={2cm 2cm 1.5cm 3cm},clip,width=0.8\linewidth]{Figures/cost.pdf}
%    \caption{Cost approximation for DSO 1.}
%    \label{fig:cost}
%\end{figure}
%\MB{should I take the cost of? Is not new from this paper.}

\section{Conclusion and Future Work} \label{sec:conclusion}
%\Ae{I think this section can be significantly shortened to gain space for explanation of the approximations and mentioned above. Just state the main conclusions of the paper, there is no need for repeating large parts of the intro. I think one sentence on future work is sufficient.}

This paper compares two approaches for reducing complexity in TSO-DSO coordination by leveraging intrinsic grid structures: flexibility aggregation via ADP and Distributed Optimization using ADMM. While ADMM approximates the optimal solution of the centralized problem closer, we achieved comparable results with ADP by using a function to approximate the aggregated flexibility cost, requiring fewer computation and communication steps. However, the quality of the ADP solution is highly dependent on the quality of the approximation of the FOR, while ADMM does not rely on approximations.

Future research will focus on improving the approximation of the flexibility cost function and deriving an inner approximation of the FOR to enhance ADP efficiency and guarantee feasibility. For distributed optimization methods like ADMM, reducing communication rounds remains a key challenge to achieving practical feasibility in grid coordination.

\renewcommand*{\bibfont}{\small}
\printbibliography

\end{document}